\documentclass[12pt]{article}
\usepackage{epsf}

\begin{document}

\thispagestyle{empty}
\renewcommand{\thefootnote}{\fnsymbol{footnote}}

\begin{flushright}
{\small
SLAC--PUB--8271\\
October 1999\\}
\end{flushright}

\vspace{.8cm}

\begin{center}
{\large \bf  QCD with SLD\footnote{Work supported by 
Department of Energy contract  DE--AC03--76SF00515.}}

\vspace{1cm}

\textbf{Toshinori Abe}\\
Stanford Linear Accelerator Center, Stanford University,
Stanford, CA  94309\\

\vspace{1cm}

\textbf{ Representing The SLD Collaboration$^{**}$}

\end{center}

\vfill

\begin{center}
{\bf Abstract}
\end{center}

\begin{quote}
We present selected new results on strong interaction physics from the SLD
experiment at the SLAC Linear Collider (SLC),
symmetry tests of $b\bar{b}g$ vertex,
the rate of secondary $b\bar{b}$ production via gluon splitting,
the $B$ hadron energy spectrum
and rapidity correlations between identified charged hadrons.
The parity violation in $Z^0 \rightarrow b\bar{b}g$ decays is
consistent with electroweak theory plus QCD.
New tests of T- and CP-conservation at the $bbg$ vertex are performed.
A new measurement of the rate of gluon splitting into $b\bar{b}$ pairs yields
$g_{b\bar{b}}=0.00307 \pm 0.00071 (stat.) \pm 0.00066 (syst.)$ 
(Preliminary).
The $B$ hadron energy spectrum is measured using a new inclusive technique,
allowing tests of predictions for its shape 
and a measurement of 
$\left< x_B \right> = 0.714 \pm 0.005 (stat.) \pm 0.007 (syst.)$ 
(Preliminary).
A study of correlations in rapidity between pairs of identified
$\pi^\pm$, $K^\pm$ and p/$\bar{\rm p}$ confirms that strangeness and baryon
number are
conserved locally, and shows local charge conservation between meson-baryon
and strange-nonstrange pairs.
Flavor-dependent long-range correlations are observed for all
combinations of these hadron species.
The first study of correlations using signed rapidities
is done and find
the first direct observation of baryon number ordering
along the $q\bar{q}$ axis.
\end{quote}

\vfill

\begin{center} 
\textit{
Presented at the XXIX International Symposium on Multiparticle Dynamics
(ISMD99)}  \\
\textit{Brown University, Providence, USA}\\
{\it August 9-13, 1999}
\end{center}

\newpage




%
\pagestyle{plain}
\section{Introduction}

The QCD physics at SLD experiment are characterized with very
excellent features of SLC/SLD.

The SLC is the only single pass $e^+e^-$ collider in the world,
running on $Z^0$ pole with highly longitudinally polarized 
electron beam and small and stable interaction point (IP).
The magnitude of the beam polarization averaged 73\%, 
providing high sensitivity to parity violating observables 
and a quark hemisphere tag of 73\% purity.
The IP can be measured to a resolution of about 4$\mu$m
in the plane transverse to the beam axis,
allowing clean $uds$, $c$ and $b$-quark event separation
with our vertex detector (VXD3)~\cite{Abe:1997bu,Abe:1999ky},
which gives remarkable impact parameter resolutions of
$\sigma_{xy} = 7.8 \oplus 33/p\sin^{3/2}\theta$ $\mu$m
and $\sigma_{rz} = 9.7 \oplus 33/p\sin^{3/2}\theta$ $\mu$m.
With topological vertex reconstruction algorithm~\cite{Jackson:1997sy},
our study put enhance on strong interaction dynamics of $b$ quark.
Efficient $\pi$, $K$ and $p$ particle identifications for
wide momentum rage are provided by the SLD Cherenkov Ring Imaging Detector 
(CRID)~\cite{Abe:1994ku},
which leads us 
a study of fragmentation process with clean $e^+e^-$ environment.

In this paper, we report on selected new QCD results at SLD experiment, 
symmetry tests of $b\bar{b}g$ vertex~\cite{Abe:1999qc},
the rate of secondary $b\bar{b}$ production 
via gluon splitting~\cite{Abe:1999qg},
the $B$ hadron energy spectrum~\cite{Abe:1999fi}
and rapidity correlations between identified charged hadrons~\cite{Abe:1999yk}.

\section{SYMMETRY TESTS IN $Z^0\rightarrow b\bar{b}g$}

In recent years, there are excitement in $Z^0 \to b\bar{b}$ sector
such as $A_b$ and $R_b$ measurements~\cite{Abbaneo:1999ub}.
It is very important to make cross check on strong interaction dynamics of
$b$ quark and through test of $b\bar{b}g$ vertex.
New tests of parity-violation in strong interactions have recently been
proposed using angular distributions in polarized 
$e^+e^- \rightarrow q\bar{q}g$ events~\cite{Burrows:1997dh}. 
In this study, 
we consider two angles, the polar angle of the quark with respect to
the electron beam direction $\theta_q$, 
and the angle between the quark-gluon and quark-electron beam planes
$\chi=\cos^{-1} (\hat{p}_q \times \hat{p}_g) \cdot 
(\hat{p}_q \times \hat{p}_e)$.
The cosine $x$ of each of these angles should be distributed as
$1+x^2+2 A_P A_Z x$, 
where the $Z^0$ polarization $A_Z = (P_e-A_e)/(1-P_eA_e)$
depends on the $e^-$ beam polarization $P_e$, 
and $A_e$ and $A_P\!=\!A_0A_q$ are
predicted by QCD plus electroweak theory.

Three-jet events (Durham algorithm, $y_{cut}=0.005$) are selected and energy
ordered.
Using 550k hadronic event sample,
the 14,658 events containing a secondary vertex with $P_T$ corrected mass 
($M_{P_T}$) above 1.5 GeV/c$^2$ in
any jet are kept, having an estimated $b\bar{b}g$ purity of 85\%.
We calculate the momentum-weighted charge of each jet $j$,
$Q_j=\Sigma_i q_i |\vec{p}_i \cdot \hat{p}_j |^{0.5}$, using the charge $q_i$
and momentum $\vec{p}_i$ of each track $i$ in the jet.
We assume that the highest-energy jet is not the gluon, and
tag it as the $b$ ($\bar{b}$) if $Q = Q_1 - Q_2 - Q_3$ is negative
(positive).
We define the $b$-quark polar angle by
$\cos\theta_b = -{\rm sign}(Q)(\hat{p}_e \cdot \hat{p}_1)$.

The left-right-forward-backward asymmetry $A_{LRFB}^b$ is shown 
as a function of $|\cos\theta_b|$ in Fig.~1.
A maximum-likelihood fit to the data yields an asymmetry parameter of
$A_P = 0.91 \pm 0.05 (stat.) \pm 0.06 (syst.)$
(Preliminary), consistent with the QCD prediction of
$A_P = 0.93 A_b = 0.87$.

We then tag one of the two lower energy jets as the gluon jet:
if jet 2 has $n_{sig}=0$ and jet 3 has $n_{sig}>0$,
where $n_{sig}$ is the number of significant tracks with normalized impact
parameter with respect to the IP $d/\sigma_d>3$,
then jet 2 is tagged as the gluon; 
otherwise jet 3 is tagged as the gluon.
The obtained result by a maximum-likelihood fit is 
$A_{\chi} = -0.014 \pm 0.035 \pm 0.002$ (Preliminary), 
to be compared with an expectation of $-0.064$.

Using these fully tagged events, we can construct observables that are
formally odd under time reversal and/or CP reversal.
For example, the triple product 
$\cos\omega^+\! = \!\vec{\sigma}_Z\! \cdot \!(\hat{p}_1 \!\times \!\hat{p}_2)$,
formed from the directions of the $Z^0$ polarization $\vec{\sigma}_Z$ and the
highest- and second highest-energy jets, is $T_N$-odd and CP-even.
Since the true time reversed experiment is not performed, this
quantity could have a nonzero $A_{LRFB}$.
A calculation~\cite{Brandenburg:1996nv}
including Standard Model final state interactions
predicts that $A_{LRFB}^{\omega^+}$ is largest for $b\bar{b}g$
events, but is only $\sim$10$^{-5}$.
The fully flavor-ordered triple product 
$\cos\omega^- \! =\! \vec{\sigma}_Z \!\cdot\!
(\hat{p}_q \!\times\! \hat{p}_{\bar{q}})$
is both $T_N$-odd and CP-odd.
Obtained $A_{LRFB}^{\omega^+}$ and $A_{LRFB}^{\omega^-}$
are consistent with zero at all $|\cos\omega|$.
Fits to the data yield 95\% C.L. limits on any $T_N$-violating and
CP-conserving or CP-violating asymmetries of $-0.038<A^+_T<0.014$ or
$-0.077<A^-_T<0.011$, respectively.

\section{The Rate of Secondary $b\bar{b}$ Production via $g \to b\bar{b}$}

The rate of secondary $b$-quark pair production 
via gluon splitting, $g \to b\bar{b}$,
is also important input for $R_b$ measurement
in $Z^0$ decays and $b$-quark production in hadron-hadron collisions.
However the rate is poorly known, both theoretically and experimentally,
despite the fact that this is one of the elementary processes in QCD.
The rate can be calculated using pQCD~\cite{Miller:1998ig},
however it is sensitive to both the $\Lambda^5_{\overline{MS}}$ parameter
and the $b$-quark mass, which results in a substantial uncertainty in
the calculation of the rate.
And the measurement is experimentally difficult due to difficulty of $B$ 
jet tag from $g\to b\bar{b}$ and large background.
Here we present a new measurement of the $g\to b\bar{b}$ rate
which takes our excellent $B$ tagging performance.

In this analysis, we use 400k hadronic event sample.
Candidate events containing a gluon splitting into a $b\bar{b}$ pair, 
$Z^0 \!\rightarrow \! q\bar{q}g \!\rightarrow \!q\bar{q}b\bar{b}$,
where the initial $q\bar{q}$ can be any flavor, are required to have 4 jets
(Durham algorithm, $y_{cut}=0.008$).
A secondary vertex is required in each of the two jets with the smallest
opening angle in the event, yielding 314 events.
This sample is dominated by background, primarily from
$Z^0 \!\rightarrow \! b\bar{b}g(g)$ events and events with a gluon splitting
into a $c\bar{c}$ pair.

A large component of the former background is $Z^0\!\rightarrow \! b\bar{b}g$
events
in which the $b$ or $\bar{b}$ jet is split into two jets by the jet finder,
and two distinct vertices from the {\it same} $B$-hadron decay are found.
Since the small beam spot allows the vertex flight directions to be
measured precisely, and the angle between the two flight directions from this
background source tends to be small, it is suppressed by a cut on this angle.

Cuts are also made
on the sum of the energies of the two jets, the angle between
the plane formed by the two selected jets and that formed by
the other two jets in the event, and the larger of the vertex masses 
($M_{P_T}$s).
The distribution of the latter quantity is
shown in Fig.~2 after all other cuts.  A clear excess of
events is visible over the expected background for masses above 2 GeV/c$^2$.
A cut at this value keeps 62 events, with an estimated background of
27.6$\pm$1.2 events.  Using this and the estimated efficiency for selecting
$g\!\rightarrow \! b\bar{b}$ splittings of 3.9\% yields a measured fraction of
hadronic events containing such a splitting of
\[
g_{b\bar{b}} = 0.00307 \pm 0.00071 \; (stat.) \pm 0.00066 \; (syst.)
\mbox{  (Preliminary).}
\]
\noindent
The systematic error is dominated by Monte Carlo statistics.
The result is consistent with and complementary to previous 
measurements~\cite{Barate:1998vs,Abreu:1997nf,Abreu:1999qh,Ref:OPAL}
;
in particular it is relatively insensitive to the modeling of the gluon
splitting process, due to the excellent efficiency for finding vertices from
low-energy $B$ hadrons.

\parbox[b]{245pt}{
\vskip 0.3in
  \parbox{220pt}{
    \epsfxsize 180pt
    \centerline{\epsfbox{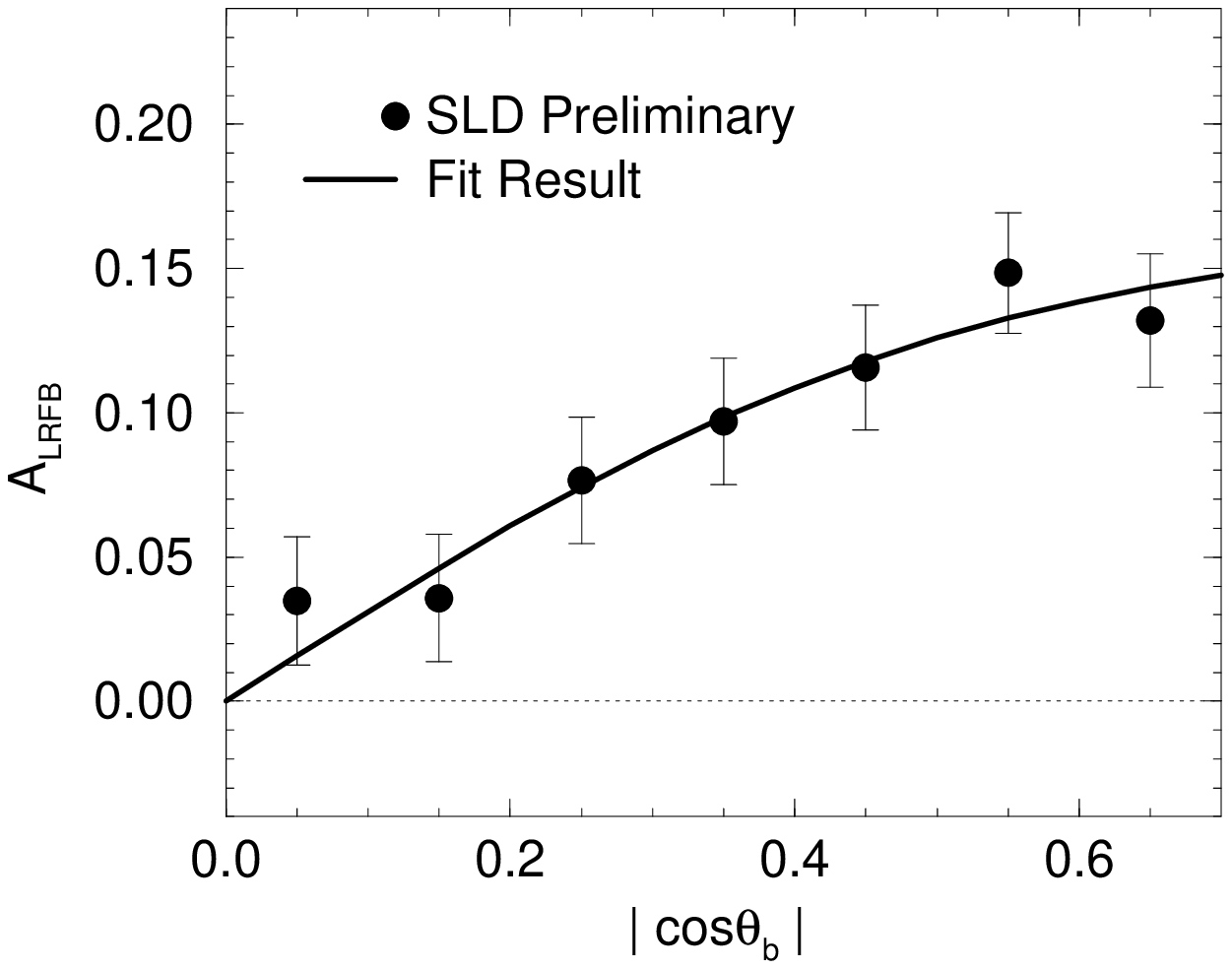}}
  }
  \vskip 0.2in
  \parbox{220pt}{
    {\footnotesize    \hspace{10pt}
      Figure 1: Left-right-forward-backward asymmetry of the $b$-quark
      polar angle in 3-jet $Z^0$ decays. The line is the result of a fit.
    }
  }
}
\parbox[b]{245pt}{
\vskip 0.3in
  \parbox{220pt}{
    \epsfxsize 180pt
    \centerline{\epsfbox{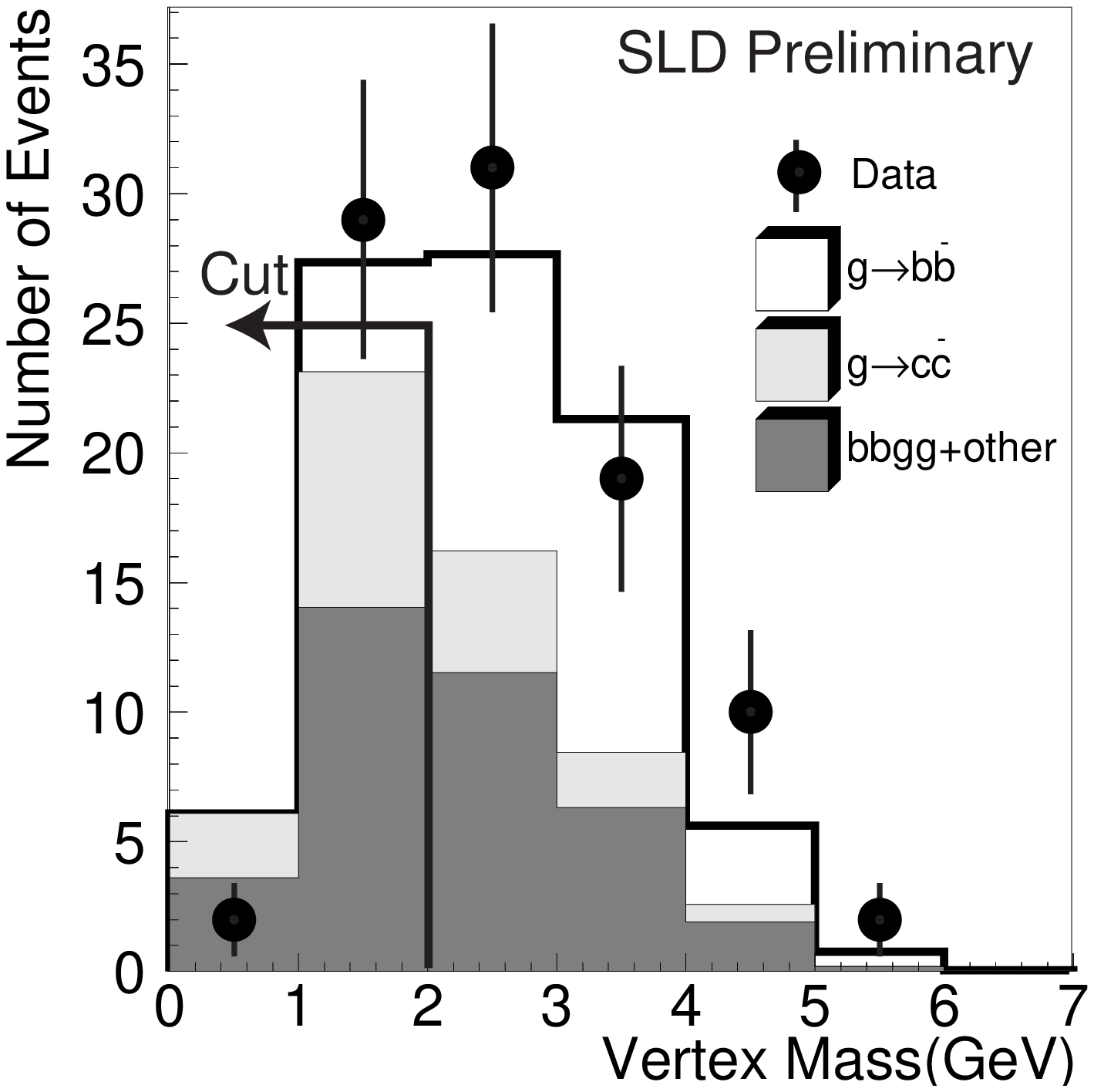}}
  }
  \vskip 0.2in
  \parbox{220pt}{
    {\footnotesize    \hspace{10pt}
      Figure 2:
      Distribution of the larger of the two vertex masses in candidate gluon 
      splitting events after all other cuts (dots).
      The backgrounds expected from the simulation are indicated.
    }
  }
}

\section{The $B$ Hadron Energy Spectrum}

Experimental studies of the $B$-hadron spectrum have been limited by the
efficiency for reconstructing the energies $E_B$ of individual $B$
hadrons with good resolution, especially for low-energy $B$ hadrons.
Here we present a new study of the $E_B$ distribution
using a novel kinematic technique and only charged tracks.
The high efficiency and good resolution for all $E_B$, results in a
measurement covering the full kinematic range.

In this analysis,
any secondary vertex in either thrust hemisphere of an event that has
$M_{p_t}>2$ GeV/c$^2$ is considered as a candidate $B$-hadron vertex.
Its flight direction is taken to be along the line joining the
IP and the vertex position.
The four-vector sum of the tracks in the vertex 
(assigned the charged pion mass)
is calculated, and the momentum component $P_t$ transverse to the flight
direction is equated with the transverse component of the ``missing" momentum.
At this point the missing mass $M_0$ and momentum along the flight direction
are still needed to determine the energy $E_B$.
Assuming a $B$ hadron mass of $M_B$ eliminates one of these unknowns,
and also allows an upper limit to be calculated on $M_0$:
\[
M^2_{0max} = M^2_B - 2M_B \sqrt{M^2_{chg} + P^2_t} + M^2_{chg},
\]
\noindent
where $M_{chg}$ is the mass of the set of tracks in the vertex.
True $M_0$ strongly tends to cluster near the $M_{0max}$.
Using $M_B=5.28$ GeV/c$^2$, equating $M^2_0$ with $M^2_{0max}$ and solving for
$E_B$ provides a good estimate of the true $E_B$.
A sample of 1938 vertices is selected from $150$k hadronic event sample, 
with an estimated $B$-hadron purity of 99.5\%.
The simulated energy resolution is 10\% on average, roughly independent
of $E_B$.

Using the obtained data distribution of the scaled energy $x_B=E_B/E_{beam}$,
we test several fragmentation models at detector level using binned $\chi^2$.
Within the context of the JETSET~\cite{Sjostrand:1994yb}
simulation,
we exclude the models of Peterson et.~al~\cite{Peterson:1983ak}
Braaten et.~al~\cite{Braaten:1995bz}, 
and Collins and Spiller~\cite{Collins:1985ms},
whereas those of the Lund group~\cite{Andersson:1983ia},
Bowler~\cite{Bowler:1981sb}, 
and Kartvelishvili~\cite{Kartvelishvili:1978pi}
are able to describe the data.
In addition, we test the UCLA~\cite{Chun:1997bh}
and HERWIG~\cite{Marchesini:1991ch} fragmentation models.
The UCLA model is consistent with the data, but that of the HERWIG model
is not.

Using four fragmentation models and four functional 
forms~\cite{Ref:adhocfunction} 
that are consistent with the data,
we unfold the data to obtain the true
distribution, concerning migration between bins.
Fig.~3 shows the unfolded $x_B$  distribution, 
where in each bin $i$ the average of the eight forms and the error bar 
includes their rms deviation due to strong simulation dependence of
the unfolding.
From these eight forms 
we extract a measurement of the mean value of the
scaled energy,
\[
\left< x_B \right> = 0.714 \pm 0.005 (stat.) \pm 0.007 (syst.) \pm 0.002 (rms)
\mbox{ (Preliminary).}
\]
\noindent
This is the most precise of the world's measurements that take the shape
dependence into account, and this uncertainty is small
since we are able to exclude a wide range of shapes.

\section{Rapidity Correlations}

Lighter identified particles are also an active field of study.
The production of strange particles and baryons is of particular interest as
they must be produced in strange-antistrange or baryon-antibaryon pairs, 
and the mechanism of their pair production can yield insights into the
fragmentation process.

Here we present detailed studies of short- and
long-range correlations between identified $\pi^\pm$, $K^\pm$ and
p/$\bar{\rm p}$.
In addition, we use the SLC beam polarization to tag the quark hemisphere in
each event and study for the first time rapidities signed such that positive
(negative) rapidity corresponds to the (anti)quark hemisphere.

For this study  we use the entire sample of hadronic events,
as well as subsamples
tagged as primary light-($uds$), $c$-, and $b$-flavor, having purities of
88\%, 39\%, and 93\%, respectively.
Charged tracks identified as $\pi^\pm$, $K^\pm$ or p/$\bar{\rm p}$ in the CRID
are considered, and their rapidities
$y=0.5\ln ((E+p_{\parallel})/(E-p_{\parallel}))$ are calculated using their
measured energies and components of momentum along the event thrust axis
$p_{\parallel}$.
For each pair of identified tracks in an event the absolute
value of the difference between their rapidities $|\Delta y|=|y_1 - y_2|$ is
considered.
Fig.~4 shows the difference between the opposite-charge and 
same-charge distributions for all six pair combinations of identified 
$\pi$, $K$ and $p$.
The excesses of the $\pi\pi$, $KK$ and $pp$ at low $|\Delta y|$ indicate
that strangeness, baryon number and electric charge conservation is local 
in the jet fragmentation process.

\parbox[b]{245pt}{
\vskip 0.3in
  \parbox{220pt}{
    \epsfxsize 180pt
    \centerline{\epsffile{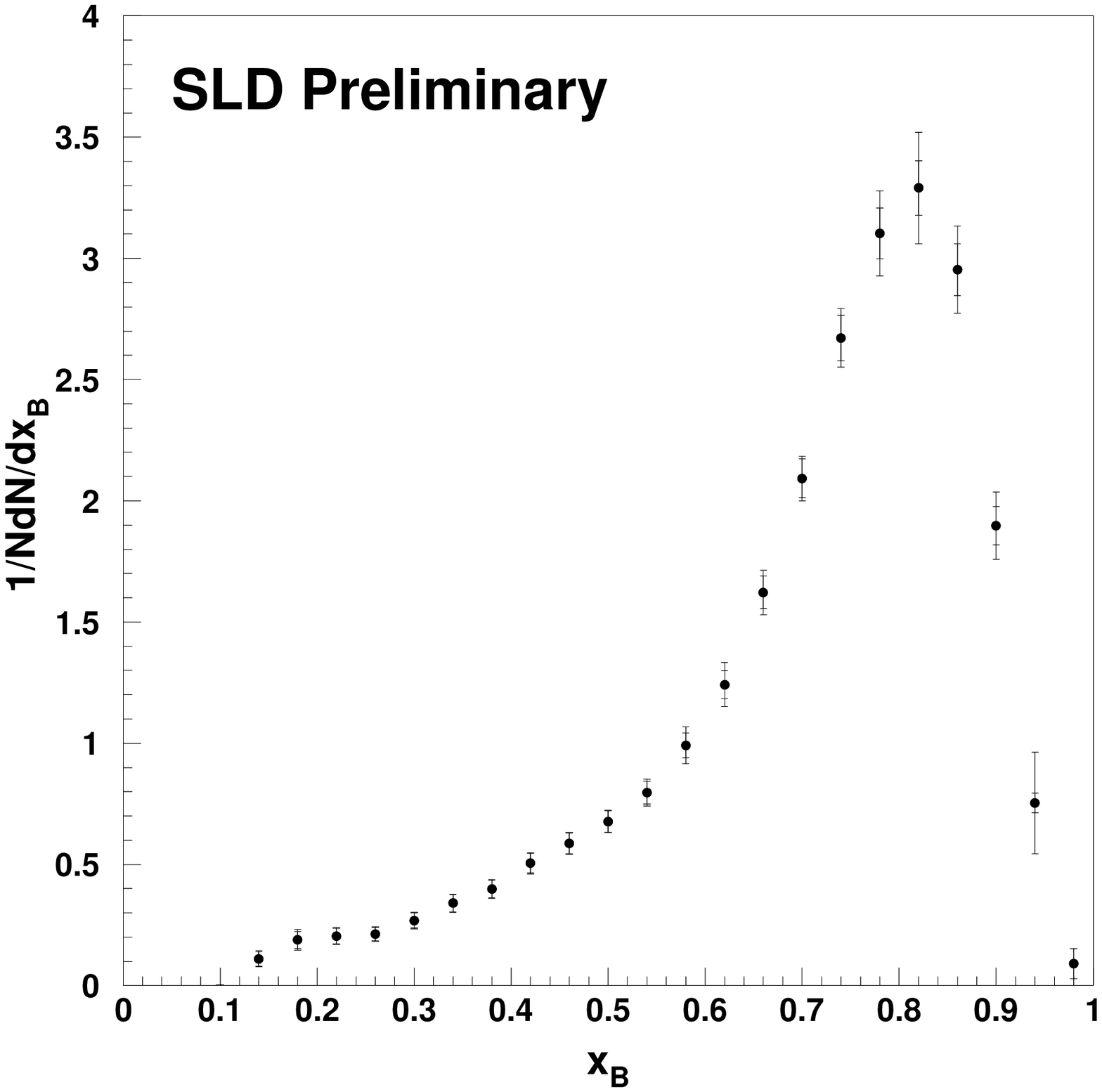}}
  }
  \vskip 0.2in
  \parbox{220pt}{
    {\footnotesize    \hspace{10pt}
      Figure 3:
      Corrected distribution of $B$-hadron energies averaged over the eight
      acceptable shapes.
      The outer error bars include the rms deviation among these shapes and 
      provide an envelope for the true shape of the distribution.
    }
  }
}
\parbox[b]{245pt}{
\vskip 0.3in
  \parbox{220pt}{
    \epsfxsize 180pt
    \centerline{\epsffile{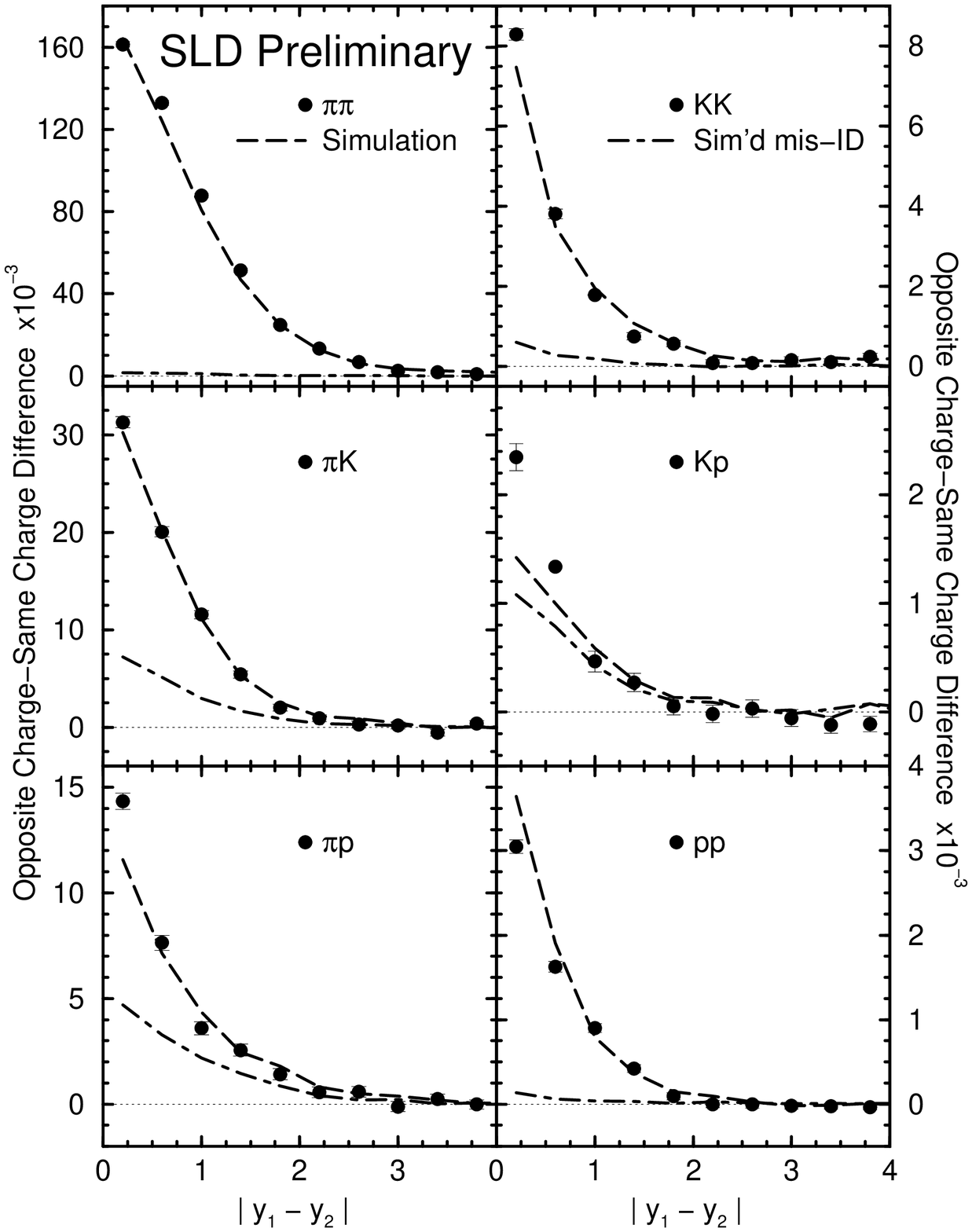}}
  }
  \vskip 0.2in
  \parbox{220pt}{
    {\footnotesize    \hspace{10pt}
      Figure 4:
      Differences between the $|\Delta y|$ distributions for opposite- and
      same-charge pairs of identified $\pi^\pm$, $K^\pm$, p/$\bar{\rm p}$ 
      in hadronic $Z^0$ decays (dots).
      The dashed (dot-dashed) lines indicate the simulated differences 
      (contributions from pairs with a misidentified track).
    }
  }
}
\vskip 0.2in

Also short-range correlations for all
three unlike pair combinations are shown in Fig.~4.
Excellent particle identification is required to observe these over the
background from $\pi\pi$ pairs in which one of the pions is misidentified.
This is the first direct observation of a fundamental feature of the jet
fragmentation process, that electric charge can be conserved locally between
a meson and a baryon, or between a strange particle and a nonstrange particle,
and suggests charge ordering along the entire fragmentation chain.

To study long-range correlations, which are expected from leading
particle production,  it is necessary to consider high momentum tracks.
The differences
between the opposite- and same-charge $|\Delta y|$ distributions for pairs of
tracks in which both have $p>9$ GeV/c are studied.
Significant correlations are observed for all pair combinations 
in light flavor events (see Fig.~5).
The JETSET model predictions are consistent with these data, 
except that no $\pi K$ correlation is predicted.

We now give the rapidity a meaningful sign by using the beam polarization to
tag the quark
hemisphere in each event, with a purity of 73\%.
The thrust axis is signed to point into this hemisphere, thus signing the
rapidity such that $y>0$ ($y<0$) in the (anti)quark hemisphere.
For pairs of hadrons one can define an ordered rapidity difference;
for hadron-antihadron pairs we define $\Delta y^{+-} = y_+ - y_-$.

Fig.~6 shows the results for $\pi^+\pi-$, $K^+K^-$ and $p\bar{p}$ pair
combinations.
The large positive difference observed for p$\bar{\rm p}$ pairs at low
$|\Delta y^{+-}|$ is the first direct observation of another fundamental
feature of jet fragmentation, namely the ordering of baryon number along the
quark-antiquark axis.
That is, the proton in a correlated
proton-antiproton pair `knows' and prefers the quark direction over the
antiquark direction.
This excess is observed at all proton momenta so cannot be attributed simply
to leading baryons.
We have searched for similar signals for strangeness and charge ordering
in the $K^+K^-$ and $\pi^+\pi^-$ samples, respectively, by isolating the
light flavors and considering a variety of momentum bins.
However no significant effects are observed.

\parbox[b]{245pt}{
\vskip 0.3in
  \parbox{220pt}{
    \epsfxsize 180pt
    \epsfysize 2.5in
    \centerline{\epsffile{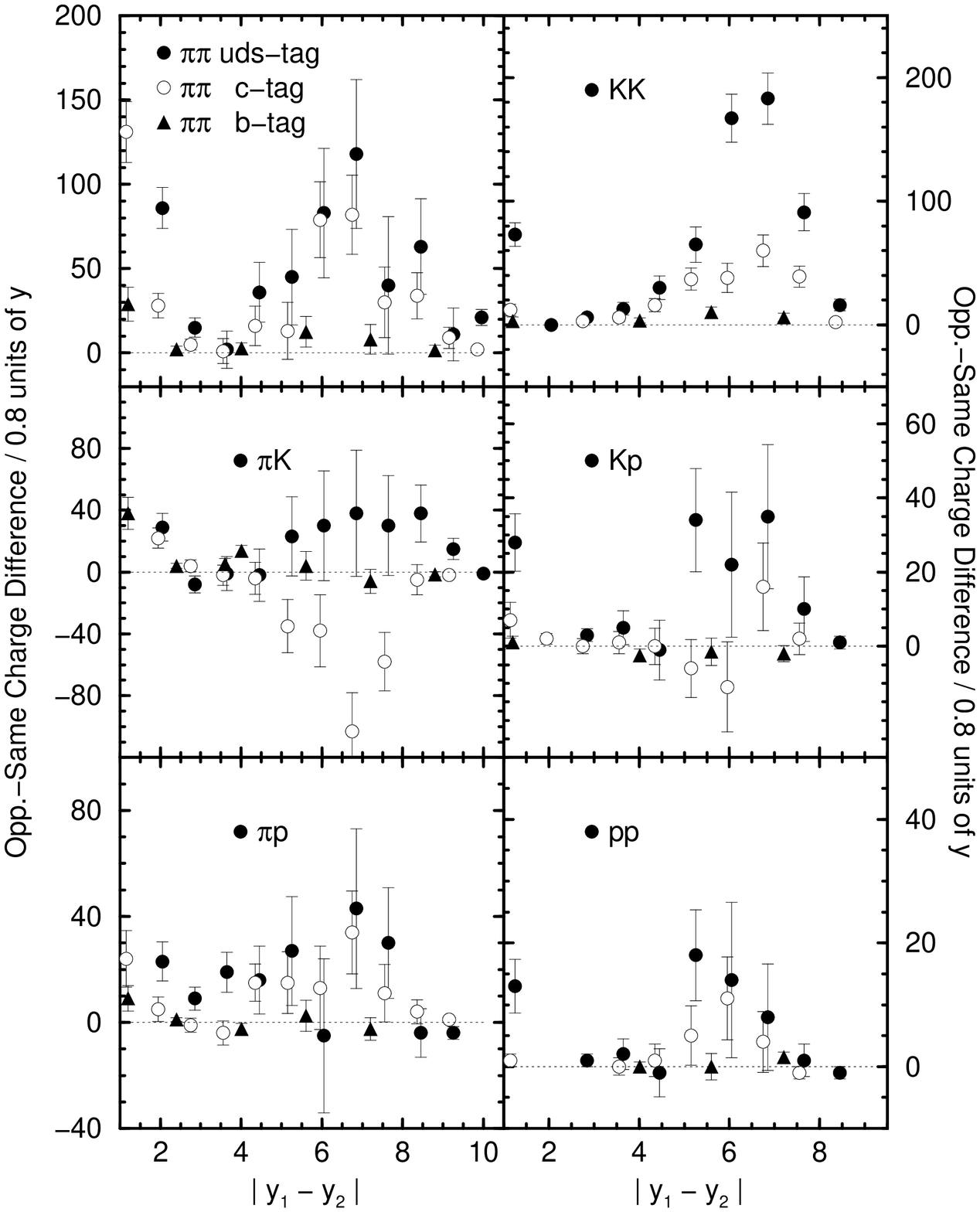}}
  }
  \vskip 0.2in
  \parbox{220pt}{
    {\footnotesize    \hspace{10pt}
      Figure 5:
      Differences between the $|\Delta y|$ distributions for 
      opposite-charge and same-charge pairs in which both tracks have 
      $p>9$ GeV/c, for the light-(dots),
      $c$-(open circles), and $b$-tagged (triangles) samples.
    }
  }
}
\parbox[b]{245pt}{
\vskip 0.3in
  \parbox{220pt}{
    \epsfxsize 180pt
    \centerline{\epsffile{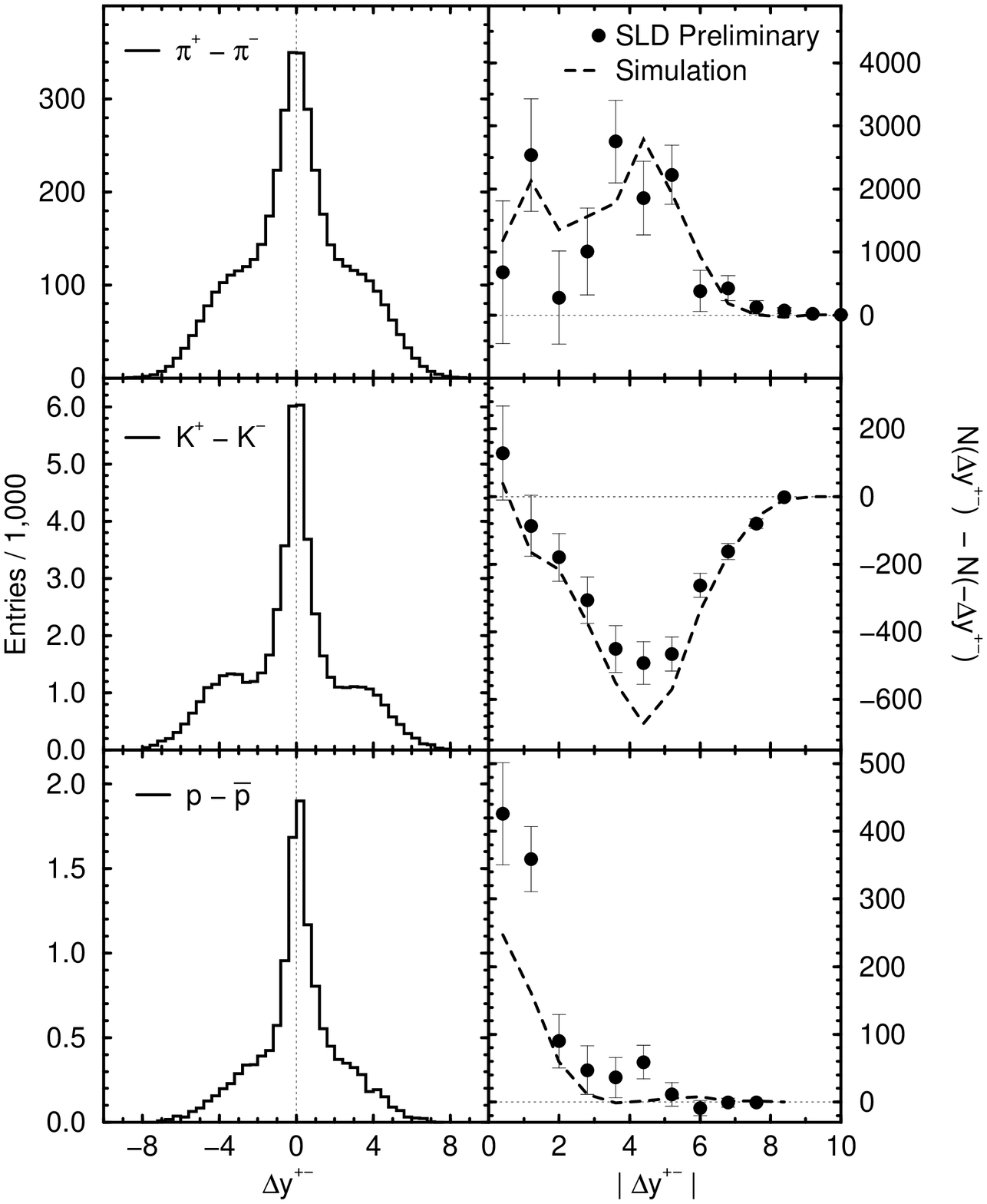}}
  }
  \vskip 0.2in
  \parbox{220pt}{
    {\footnotesize    \hspace{10pt}
      Figure 6:
      Distributions (left) of the ordered signed rapidity difference 
      $\Delta y^{+-}$, and differences (right) between the positive 
      and negative sides of each distribution.
      The dashed lines indicate the predictions of the JETSET simulation.
    }
  }
}

\section{Summary}

We use the excellent SLD vertexing and particle identification, and
the high SLC $e^-$ beam polarization to make several new tests of QCD.
Using 3-jet final states in which jets are tagged as $b$,
$\bar{b}$ or gluon jets:
we find the parity violation in $Z^0\!\rightarrow \! b\bar{b}g$ decays to be
consistent with electroweak theory plus QCD;
and we perform new tests of T- and CP-conservation in strong interactions.
Using 4-jet final states in which the two most collinear jets are tagged as
$b$/$\bar{b}$, we measure the rate of gluon splitting into a $b\bar{b}$ pair in
hadronic $Z^0$ decays, 
$g_{b\bar{b}} = 0.00307 \pm 0.00071 (stat.) \pm 0.00066 (syst.)$ 
(Preliminary).

A new inclusive technique for measuring the energies of individual $B$ hadrons
is studied.
We exclude several fragmentation functions, constrain the shape,
and obtain an averaged scaled energy of 
$\left< x_B \right> = 0.714 \pm 0.005 (stat.) \pm 0.007 (syst.) \pm 0.002
 (shape)$ (Preliminary).

Considering pairs of identified $\pi^\pm$, $K^\pm$ and p/$\bar{\rm p}$,
we confirm that the conservation of
strangeness, baryon number and electric charge
quantum numbers
is local in the jet fragmentation process, 
and observe local charge conservation
between mesons and baryons and between strange and nonstrange particles.
Long range correlations are also observed between all these pair combinations.
The first study of ordered correlations in signed rapidity provides
additional new information on fragmentation, including the first
direct observation of baryon number ordering along the quark-antiquark axis.


\section*{**List of Authors}

%
%
%
\begin{center}
\def\iADEL{$^{(1)}$}
\def\iAOMORI{$^{(2)}$}
\def\iBOLO{$^{(3)}$}
\def\iBRI{$^{(4)}$}
\def\iBRUN{$^{(5)}$}
\def\iBU{$^{(6)}$}
\def\iCINC{$^{(7)}$}
\def\iCOLO{$^{(8)}$}
\def\iCOLU{$^{(9)}$}
\def\iCSU{$^{(10)}$}
\def\iFERR{$^{(11)}$}
\def\iFRAS{$^{(12)}$}
\def\iILLI{$^{(13)}$}
\def\iJHU{$^{(14)}$}
\def\iLBL{$^{(15)}$}
\def\iLTU{$^{(16)}$}
\def\iMASS{$^{(17)}$}
\def\iMISSI{$^{(18)}$}
\def\iMIT{$^{(19)}$}
\def\iMOSCOW{$^{(20)}$}
\def\iNAGO{$^{(21)}$}
\def\iOREG{$^{(22)}$}
\def\iOXF{$^{(23)}$}
\def\iPADO{$^{(24)}$}
\def\iPERU{$^{(25)}$}
\def\iPISA{$^{(26)}$}
\def\iRAL{$^{(27)}$}
\def\iRUTG{$^{(28)}$}
\def\iSLAC{$^{(29)}$}
\def\iSOGA{$^{(30)}$}
\def\iSOONG{$^{(31)}$}
\def\iTENN{$^{(32)}$}
\def\iTOHO{$^{(33)}$}
\def\iUCSB{$^{(34)}$}
\def\iUCSC{$^{(35)}$}
\def\iUVIC{$^{(36)}$}
\def\iVAND{$^{(37)}$}
\def\iWASH{$^{(38)}$}
\def\iWISC{$^{(39)}$}
\def\iYALE{$^{(40)}$}

  \baselineskip=.75\baselineskip  
\mbox{Kenji  Abe\unskip,\iNAGO}
\mbox{Koya Abe\unskip,\iTOHO}
\mbox{T. Abe\unskip,\iSLAC}
\mbox{I. Adam\unskip,\iSLAC}
\mbox{T.  Akagi\unskip,\iSLAC}
\mbox{H. Akimoto\unskip,\iSLAC}
\mbox{N.J. Allen\unskip,\iBRUN}
\mbox{W.W. Ash\unskip,\iSLAC}
\mbox{D. Aston\unskip,\iSLAC}
\mbox{K.G. Baird\unskip,\iMASS}
\mbox{C. Baltay\unskip,\iYALE}
\mbox{H.R. Band\unskip,\iWISC}
\mbox{M.B. Barakat\unskip,\iLTU}
\mbox{O. Bardon\unskip,\iMIT}
\mbox{T.L. Barklow\unskip,\iSLAC}
\mbox{G.L. Bashindzhagyan\unskip,\iMOSCOW}
\mbox{J.M. Bauer\unskip,\iMISSI}
\mbox{G. Bellodi\unskip,\iOXF}
\mbox{A.C. Benvenuti\unskip,\iBOLO}
\mbox{G.M. Bilei\unskip,\iPERU}
\mbox{D. Bisello\unskip,\iPADO}
\mbox{G. Blaylock\unskip,\iMASS}
\mbox{J.R. Bogart\unskip,\iSLAC}
\mbox{G.R. Bower\unskip,\iSLAC}
\mbox{J.E. Brau\unskip,\iOREG}
\mbox{M. Breidenbach\unskip,\iSLAC}
\mbox{W.M. Bugg\unskip,\iTENN}
\mbox{D. Burke\unskip,\iSLAC}
\mbox{T.H. Burnett\unskip,\iWASH}
\mbox{P.N. Burrows\unskip,\iOXF}
\mbox{R.M. Byrne\unskip,\iMIT}
\mbox{A. Calcaterra\unskip,\iFRAS}
\mbox{D. Calloway\unskip,\iSLAC}
\mbox{B. Camanzi\unskip,\iFERR}
\mbox{M. Carpinelli\unskip,\iPISA}
\mbox{R. Cassell\unskip,\iSLAC}
\mbox{R. Castaldi\unskip,\iPISA}
\mbox{A. Castro\unskip,\iPADO}
\mbox{M. Cavalli-Sforza\unskip,\iUCSC}
\mbox{A. Chou\unskip,\iSLAC}
\mbox{E. Church\unskip,\iWASH}
\mbox{H.O. Cohn\unskip,\iTENN}
\mbox{J.A. Coller\unskip,\iBU}
\mbox{M.R. Convery\unskip,\iSLAC}
\mbox{V. Cook\unskip,\iWASH}
\mbox{R.F. Cowan\unskip,\iMIT}
\mbox{D.G. Coyne\unskip,\iUCSC}
\mbox{G. Crawford\unskip,\iSLAC}
\mbox{C.J.S. Damerell\unskip,\iRAL}
\mbox{M.N. Danielson\unskip,\iCOLO}
\mbox{M. Daoudi\unskip,\iSLAC}
\mbox{N. de Groot\unskip,\iBRI}
\mbox{R. Dell'Orso\unskip,\iPERU}
\mbox{P.J. Dervan\unskip,\iBRUN}
\mbox{R. de Sangro\unskip,\iFRAS}
\mbox{M. Dima\unskip,\iCSU}
\mbox{D.N. Dong\unskip,\iMIT}
\mbox{M. Doser\unskip,\iSLAC}
\mbox{R. Dubois\unskip,\iSLAC}
\mbox{B.I. Eisenstein\unskip,\iILLI}
\mbox{I.Erofeeva\unskip,\iMOSCOW}
\mbox{V. Eschenburg\unskip,\iMISSI}
\mbox{E. Etzion\unskip,\iWISC}
\mbox{S. Fahey\unskip,\iCOLO}
\mbox{D. Falciai\unskip,\iFRAS}
\mbox{C. Fan\unskip,\iCOLO}
\mbox{J.P. Fernandez\unskip,\iUCSC}
\mbox{M.J. Fero\unskip,\iMIT}
\mbox{K. Flood\unskip,\iMASS}
\mbox{R. Frey\unskip,\iOREG}
\mbox{J. Gifford\unskip,\iUVIC}
\mbox{T. Gillman\unskip,\iRAL}
\mbox{G. Gladding\unskip,\iILLI}
\mbox{S. Gonzalez\unskip,\iMIT}
\mbox{E.R. Goodman\unskip,\iCOLO}
\mbox{E.L. Hart\unskip,\iTENN}
\mbox{J.L. Harton\unskip,\iCSU}
\mbox{K. Hasuko\unskip,\iTOHO}
\mbox{S.J. Hedges\unskip,\iBU}
\mbox{S.S. Hertzbach\unskip,\iMASS}
\mbox{M.D. Hildreth\unskip,\iSLAC}
\mbox{J. Huber\unskip,\iOREG}
\mbox{M.E. Huffer\unskip,\iSLAC}
\mbox{E.W. Hughes\unskip,\iSLAC}
\mbox{X. Huynh\unskip,\iSLAC}
\mbox{H. Hwang\unskip,\iOREG}
\mbox{M. Iwasaki\unskip,\iOREG}
\mbox{D.J. Jackson\unskip,\iRAL}
\mbox{P. Jacques\unskip,\iRUTG}
\mbox{J.A. Jaros\unskip,\iSLAC}
\mbox{Z.Y. Jiang\unskip,\iSLAC}
\mbox{A.S. Johnson\unskip,\iSLAC}
\mbox{J.R. Johnson\unskip,\iWISC}
\mbox{R.A. Johnson\unskip,\iCINC}
\mbox{T. Junk\unskip,\iSLAC}
\mbox{R. Kajikawa\unskip,\iNAGO}
\mbox{M. Kalelkar\unskip,\iRUTG}
\mbox{Y. Kamyshkov\unskip,\iTENN}
\mbox{H.J. Kang\unskip,\iRUTG}
\mbox{I. Karliner\unskip,\iILLI}
\mbox{H. Kawahara\unskip,\iSLAC}
\mbox{Y.D. Kim\unskip,\iSOGA}
\mbox{M.E. King\unskip,\iSLAC}
\mbox{R. King\unskip,\iSLAC}
\mbox{R.R. Kofler\unskip,\iMASS}
\mbox{N.M. Krishna\unskip,\iCOLO}
\mbox{R.S. Kroeger\unskip,\iMISSI}
\mbox{M. Langston\unskip,\iOREG}
\mbox{A. Lath\unskip,\iMIT}
\mbox{D.W.G. Leith\unskip,\iSLAC}
\mbox{V. Lia\unskip,\iMIT}
\mbox{C.Lin\unskip,\iMASS}
\mbox{M.X. Liu\unskip,\iYALE}
\mbox{X. Liu\unskip,\iUCSC}
\mbox{M. Loreti\unskip,\iPADO}
\mbox{A. Lu\unskip,\iUCSB}
\mbox{H.L. Lynch\unskip,\iSLAC}
\mbox{J. Ma\unskip,\iWASH}
\mbox{M. Mahjouri\unskip,\iMIT}
\mbox{G. Mancinelli\unskip,\iRUTG}
\mbox{S. Manly\unskip,\iYALE}
\mbox{G. Mantovani\unskip,\iPERU}
\mbox{T.W. Markiewicz\unskip,\iSLAC}
\mbox{T. Maruyama\unskip,\iSLAC}
\mbox{H. Masuda\unskip,\iSLAC}
\mbox{E. Mazzucato\unskip,\iFERR}
\mbox{A.K. McKemey\unskip,\iBRUN}
\mbox{B.T. Meadows\unskip,\iCINC}
\mbox{G. Menegatti\unskip,\iFERR}
\mbox{R. Messner\unskip,\iSLAC}
\mbox{P.M. Mockett\unskip,\iWASH}
\mbox{K.C. Moffeit\unskip,\iSLAC}
\mbox{T.B. Moore\unskip,\iYALE}
\mbox{M.Morii\unskip,\iSLAC}
\mbox{D. Muller\unskip,\iSLAC}
\mbox{V. Murzin\unskip,\iMOSCOW}
\mbox{T. Nagamine\unskip,\iTOHO}
\mbox{S. Narita\unskip,\iTOHO}
\mbox{U. Nauenberg\unskip,\iCOLO}
\mbox{H. Neal\unskip,\iSLAC}
\mbox{M. Nussbaum\unskip,\iCINC}
\mbox{N. Oishi\unskip,\iNAGO}
\mbox{D. Onoprienko\unskip,\iTENN}
\mbox{L.S. Osborne\unskip,\iMIT}
\mbox{R.S. Panvini\unskip,\iVAND}
\mbox{C.H. Park\unskip,\iSOONG}
\mbox{T.J. Pavel\unskip,\iSLAC}
\mbox{I. Peruzzi\unskip,\iFRAS}
\mbox{M. Piccolo\unskip,\iFRAS}
\mbox{L. Piemontese\unskip,\iFERR}
\mbox{K.T. Pitts\unskip,\iOREG}
\mbox{R.J. Plano\unskip,\iRUTG}
\mbox{R. Prepost\unskip,\iWISC}
\mbox{C.Y. Prescott\unskip,\iSLAC}
\mbox{G.D. Punkar\unskip,\iSLAC}
\mbox{J. Quigley\unskip,\iMIT}
\mbox{B.N. Ratcliff\unskip,\iSLAC}
\mbox{T.W. Reeves\unskip,\iVAND}
\mbox{J. Reidy\unskip,\iMISSI}
\mbox{P.L. Reinertsen\unskip,\iUCSC}
\mbox{P.E. Rensing\unskip,\iSLAC}
\mbox{L.S. Rochester\unskip,\iSLAC}
\mbox{P.C. Rowson\unskip,\iCOLU}
\mbox{J.J. Russell\unskip,\iSLAC}
\mbox{O.H. Saxton\unskip,\iSLAC}
\mbox{T. Schalk\unskip,\iUCSC}
\mbox{R.H. Schindler\unskip,\iSLAC}
\mbox{B.A. Schumm\unskip,\iUCSC}
\mbox{J. Schwiening\unskip,\iSLAC}
\mbox{S. Sen\unskip,\iYALE}
\mbox{V.V. Serbo\unskip,\iSLAC}
\mbox{M.H. Shaevitz\unskip,\iCOLU}
\mbox{J.T. Shank\unskip,\iBU}
\mbox{G. Shapiro\unskip,\iLBL}
\mbox{D.J. Sherden\unskip,\iSLAC}
\mbox{K.D. Shmakov\unskip,\iTENN}
\mbox{C. Simopoulos\unskip,\iSLAC}
\mbox{N.B. Sinev\unskip,\iOREG}
\mbox{S.R. Smith\unskip,\iSLAC}
\mbox{M.B. Smy\unskip,\iCSU}
\mbox{J.A. Snyder\unskip,\iYALE}
\mbox{H. Staengle\unskip,\iCSU}
\mbox{A. Stahl\unskip,\iSLAC}
\mbox{P. Stamer\unskip,\iRUTG}
\mbox{H. Steiner\unskip,\iLBL}
\mbox{R. Steiner\unskip,\iADEL}
\mbox{M.G. Strauss\unskip,\iMASS}
\mbox{D. Su\unskip,\iSLAC}
\mbox{F. Suekane\unskip,\iTOHO}
\mbox{A. Sugiyama\unskip,\iNAGO}
\mbox{S. Suzuki\unskip,\iNAGO}
\mbox{M. Swartz\unskip,\iJHU}
\mbox{A. Szumilo\unskip,\iWASH}
\mbox{T. Takahashi\unskip,\iSLAC}
\mbox{F.E. Taylor\unskip,\iMIT}
\mbox{J. Thom\unskip,\iSLAC}
\mbox{E. Torrence\unskip,\iMIT}
\mbox{N.K. Toumbas\unskip,\iSLAC}
\mbox{T. Usher\unskip,\iSLAC}
\mbox{C. Vannini\unskip,\iPISA}
\mbox{J. Va'vra\unskip,\iSLAC}
\mbox{E. Vella\unskip,\iSLAC}
\mbox{J.P. Venuti\unskip,\iVAND}
\mbox{R. Verdier\unskip,\iMIT}
\mbox{P.G. Verdini\unskip,\iPISA}
\mbox{D.L. Wagner\unskip,\iCOLO}
\mbox{S.R. Wagner\unskip,\iSLAC}
\mbox{A.P. Waite\unskip,\iSLAC}
\mbox{S. Walston\unskip,\iOREG}
\mbox{S.J. Watts\unskip,\iBRUN}
\mbox{A.W. Weidemann\unskip,\iTENN}
\mbox{E. R. Weiss\unskip,\iWASH}
\mbox{J.S. Whitaker\unskip,\iBU}
\mbox{S.L. White\unskip,\iTENN}
\mbox{F.J. Wickens\unskip,\iRAL}
\mbox{B. Williams\unskip,\iCOLO}
\mbox{D.C. Williams\unskip,\iMIT}
\mbox{S.H. Williams\unskip,\iSLAC}
\mbox{S. Willocq\unskip,\iMASS}
\mbox{R.J. Wilson\unskip,\iCSU}
\mbox{W.J. Wisniewski\unskip,\iSLAC}
\mbox{J. L. Wittlin\unskip,\iMASS}
\mbox{M. Woods\unskip,\iSLAC}
\mbox{G.B. Word\unskip,\iVAND}
\mbox{T.R. Wright\unskip,\iWISC}
\mbox{J. Wyss\unskip,\iPADO}
\mbox{R.K. Yamamoto\unskip,\iMIT}
\mbox{J.M. Yamartino\unskip,\iMIT}
\mbox{X. Yang\unskip,\iOREG}
\mbox{J. Yashima\unskip,\iTOHO}
\mbox{S.J. Yellin\unskip,\iUCSB}
\mbox{C.C. Young\unskip,\iSLAC}
\mbox{H. Yuta\unskip,\iAOMORI}
\mbox{G. Zapalac\unskip,\iWISC}
\mbox{R.W. Zdarko\unskip,\iSLAC}
\mbox{J. Zhou\unskip.\iOREG}

\it
  \vskip \baselineskip                   
  \centerline{(The SLD Collaboration)}   
  \vskip \baselineskip        
  \baselineskip=.75\baselineskip   
\iADEL
  Adelphi University, Garden City, New York 11530, \break
\iAOMORI
  Aomori University, Aomori , 030 Japan, \break
\iBOLO
  INFN Sezione di Bologna, I-40126, Bologna, Italy, \break
\iBRI
  University of Bristol, Bristol, U.K., \break
\iBRUN
  Brunel University, Uxbridge, Middlesex, UB8 3PH United Kingdom, \break
\iBU
  Boston University, Boston, Massachusetts 02215, \break
\iCINC
  University of Cincinnati, Cincinnati, Ohio 45221, \break
\iCOLO
  University of Colorado, Boulder, Colorado 80309, \break
\iCOLU
  Columbia University, New York, New York 10533, \break
\iCSU
  Colorado State University, Ft. Collins, Colorado 80523, \break
\iFERR
  INFN Sezione di Ferrara and Universita di Ferrara, I-44100 Ferrara, Italy, \break
\iFRAS
  INFN Lab. Nazionali di Frascati, I-00044 Frascati, Italy, \break
\iILLI
  University of Illinois, Urbana, Illinois 61801, \break
\iJHU
  Johns Hopkins University,  Baltimore, Maryland 21218-2686, \break
\iLBL
  Lawrence Berkeley Laboratory, University of California, Berkeley, California 94720, \break
\iLTU
  Louisiana Technical University, Ruston,Louisiana 71272, \break
\iMASS
  University of Massachusetts, Amherst, Massachusetts 01003, \break
\iMISSI
  University of Mississippi, University, Mississippi 38677, \break
\iMIT
  Massachusetts Institute of Technology, Cambridge, Massachusetts 02139, \break
\iMOSCOW
  Institute of Nuclear Physics, Moscow State University, 119899, Moscow Russia, \break
\iNAGO
  Nagoya University, Chikusa-ku, Nagoya, 464 Japan, \break
\iOREG
  University of Oregon, Eugene, Oregon 97403, \break
\iOXF
  Oxford University, Oxford, OX1 3RH, United Kingdom, \break
\iPADO
  INFN Sezione di Padova and Universita di Padova I-35100, Padova, Italy, \break
\iPERU
  INFN Sezione di Perugia and Universita di Perugia, I-06100 Perugia, Italy, \break
\iPISA
  INFN Sezione di Pisa and Universita di Pisa, I-56010 Pisa, Italy, \break
\iRAL
  Rutherford Appleton Laboratory, Chilton, Didcot, Oxon OX11 0QX United Kingdom, \break
\iRUTG
  Rutgers University, Piscataway, New Jersey 08855, \break
\iSLAC
  Stanford Linear Accelerator Center, Stanford University, Stanford, California 94309, \break
\iSOGA
  Sogang University, Seoul, Korea, \break
\iSOONG
  Soongsil University, Seoul, Korea 156-743, \break
\iTENN
  University of Tennessee, Knoxville, Tennessee 37996, \break
\iTOHO
  Tohoku University, Sendai 980, Japan, \break
\iUCSB
  University of California at Santa Barbara, Santa Barbara, California 93106, \break
\iUCSC
  University of California at Santa Cruz, Santa Cruz, California 95064, \break
\iUVIC
  University of Victoria, Victoria, British Columbia, Canada V8W 3P6, \break
\iVAND
  Vanderbilt University, Nashville,Tennessee 37235, \break
\iWASH
  University of Washington, Seattle, Washington 98105, \break
\iWISC
  University of Wisconsin, Madison,Wisconsin 53706, \break
\iYALE
  Yale University, New Haven, Connecticut 06511. \break

\rm
%

\end{center}


\begin{thebibliography}{99}
\bibitem{Ref:adhocfunction}
We also test several ad hoc functional forms $f(x_B,\vec{\lambda})$ of the
observable $x_B$ itself, by minimizing $\chi^2$ with respect to the
parameter(s) $\vec{\lambda}$.
Four functions, the Peterson function, two generalizations thereof, 
and a sixth order polynomial, are found to describe the data.
\end{thebibliography}
\end{document}